\documentclass[twocolumn,superscriptaddress]{revtex4-1}
\usepackage{amsmath}
\usepackage{graphicx}
\usepackage{varwidth}
\usepackage{algorithm}
\usepackage[noend]{algpseudocode}
\usepackage{color}
\usepackage{units}
\usepackage{hyperref}
\usepackage{mathtools}

\newcommand\floor[1]{\lfloor#1\rfloor}

\newcommand{\RA}[1]{{\color{blue}#1}}

\newcommand{\bs}{\mathbf{s}}

\newcommand{\bv}{\mathbf{v}}

\usepackage{float}

\newcommand{\beginsupplement}{%
        \setcounter{table}{0}
        \renewcommand{\thetable}{S\arabic{table}}%
        \setcounter{figure}{0}
        \renewcommand{\thefigure}{S\arabic{figure}}%
     }

\makeatletter
\def\@seccntformat#1{%
  \expandafter\ifx\csname c@#1\endcsname\c@section\else
  \csname the#1\endcsname\quad
  \fi}
\makeatother

\begin{document}

\title{Physical epistatic landscape of antibody binding affinity}

\author{Rhys M. Adams}
\thanks{Current address: Francis Crick Institute, 1 Midland Rd, London NW1 1AT, United Kingdom.}
\affiliation{Laboratoire de Physique Th\'eorique, CNRS, UPMC
  (Sorbonne University), and \'Ecole Normale Sup\'erieure (PSL), 24, rue Lhomond, 75005 Paris, France}
\affiliation{Simons Center for Quantitative Biology, Cold Spring
  Harbor Laboratory, 1 Bungtown Rd., Cold Spring Harbor, NY, 11724,
  USA}
\author{Justin B. Kinney}
\affiliation{Simons Center for Quantitative Biology, Cold Spring
  Harbor Laboratory, 1 Bungtown Rd., Cold Spring Harbor, NY, 11724,
  USA}
\author{Aleksandra M. Walczak}
\thanks{Corresponding authors: \url{awalczak@lpt.ens.fr}, \url{tmora@lps.ens.fr}}
\affiliation{Laboratoire de Physique Th\'eorique, CNRS, UPMC
  (Sorbonne University), and \'Ecole Normale Sup\'erieure (PSL), 24,
  rue Lhomond, 75005 Paris, France}
  \author{Thierry Mora}
\thanks{Corresponding authors: \url{awalczak@lpt.ens.fr}, \url{tmora@lps.ens.fr}}
\affiliation{Laboratoire de Physique Statistique, CNRS, UPMC
  (Sorbonne University), Paris-Diderot University, and \'Ecole Normale Sup\'erieure (PSL), 24, rue Lhomond, 75005 Paris, France}

\date{\today}

\begin{abstract}
 % abstract
Affinity maturation produces antibodies that bind antigens with high specificity by accumulating mutations in the antibody sequence. Mapping out the antibody-antigen affinity landscape can give us insight into the accessible paths during this rapid evolutionary process. By developing a carefully controlled null model for noninteracting mutations, we characterized epistasis in affinity measurements of a large library of antibody variants obtained by Tite-Seq, a recently introduced Deep Mutational Scan method yielding physical values of the binding constant. We show that representing affinity as the binding free energy minimizes epistasis. Yet, we find that epistatically interacting sites contribute substantially to binding. In addition to negative epistasis, we report a large amount of beneficial epistasis, enlarging the space of high-affinity antibodies as well as their mutational accessibility.  These properties suggest that the degeneracy of antibody sequences that can bind a given antigen is enhanced by epistasis --- an important property for vaccine design.

\end{abstract}

\maketitle

 % introduction
To ensure a reliable response and to neutralize foreign pathogens, the adaptive immune system relies on affinity maturation. In this process, antibody receptors expressed by B cells undergo an accelerated Darwinian evolution through random mutations and selection for affinity against foreign epitopes \cite{Cobey2015}. Mature antibodies can accumulate up to 20\% hypermutations, leading to up to a 10,000 fold improvement in binding affinity \cite{Eisen1964}. Affinity maturation also produces broadly neutralizing antibodies that target conserved regions of the pathogen, of particular importance for vaccine design against fast evolving viruses \cite{Corti2013}. Despite extensive experimental and theoretical work, the key determinants of antibody specificity and evolvability are still poorly understood, mainly because the sequence-to-affinity relationship is difficult to measure comprehensively or to predict computationally \cite{Esmaielbeiki2016}.

A major confounding factor in characterizing the sequence dependence of any protein function, including affinity, is the pervasiveness of epistasis, the phenomenon by which different mutations interact with each other \cite{Phillips2008}. Theory \cite{carter_role_2005,Good2015,paixao_effect_2016} and genomic data \cite{Breen2012} suggest that inter- and intragenic epistasis plays a major role in molecular evolution, by constraining the set of accessible evolutionary trajectories towards adapted phenotypes \cite{weinreich_darwinian_2006,Poelwijk2007,Gong2013,Anderson2015,podgornaia_protein_2015}, enhancing evolvability through stabilizing mutations \cite{Bloom2006,Bloom2010}, or slowing down adaptation by the law of diminishing returns \cite{chou_diminishing_2011,kryazhimskiy_global_2014}. Evidence for epistasis in antibody affinity include direct observations of cooperativity between mutations \cite{Midelfort:2004iq,koenig_deep_2015}, the dependence of mutational effects on sequence background \cite{Boyer2016}, and statistical co-variation of residues in large sequence datasets \cite{mora_maximum_2010,Asti2016}.

Intragenic epistasis has mostly been studied either by measuring the fitness of all possible mutational intermediates between two variants \cite{weinreich_darwinian_2006,Schenk2013,Szendro2013,de_visser_empirical_2014,Sarkisyan2016}, or by comparing the effect of mutations in different backgrounds \cite{Jacquier2013,Bank2015,Boyer2016}.
Many such studies rely on a particular measure of fitness rather than a well-defined physical phenotype.
{Deep mutational scans (DMS) \cite{Fowler2014} can comprehensively map out the epistatic landscape of many genetic variants \cite{Araya:2012p13011,olson_comprehensive_2014,podgornaia_protein_2015}.}
However, most DMS methods rely on noisy selection and do not measure the biophysical quantity of interest directly \cite{vodnik_phage_2011}, introducing both nonlinearities and noise that could be misinterpreted as epistasis.

Here we analyze the detailed epistatic landscape of an antibody's binding free energy to its cognate antigen (the 4-4-20 antibody fragment against fluorescein), using data previously obtained by Tite-Seq, a recently introduced DMS variant that 
accurately measures protein binding affinity in physical units of molarity \cite{adams_measuring_2016}. By comparing to a simple additive model of mutations on the binding free energy, and carefully controling for measurement noise and nonlinearities, we find that epistasis significantly contributes to the antibody's affinity. This epistasis is not uniformly distributed, but instead favors certain residue pairs across the protein. We use our results to analyze how epistasis both constrains and enlarges the set of possible evolutionary paths leading to high-affinity sequences.

 % results
\begin{figure*}
\includegraphics[width=0.65\linewidth]{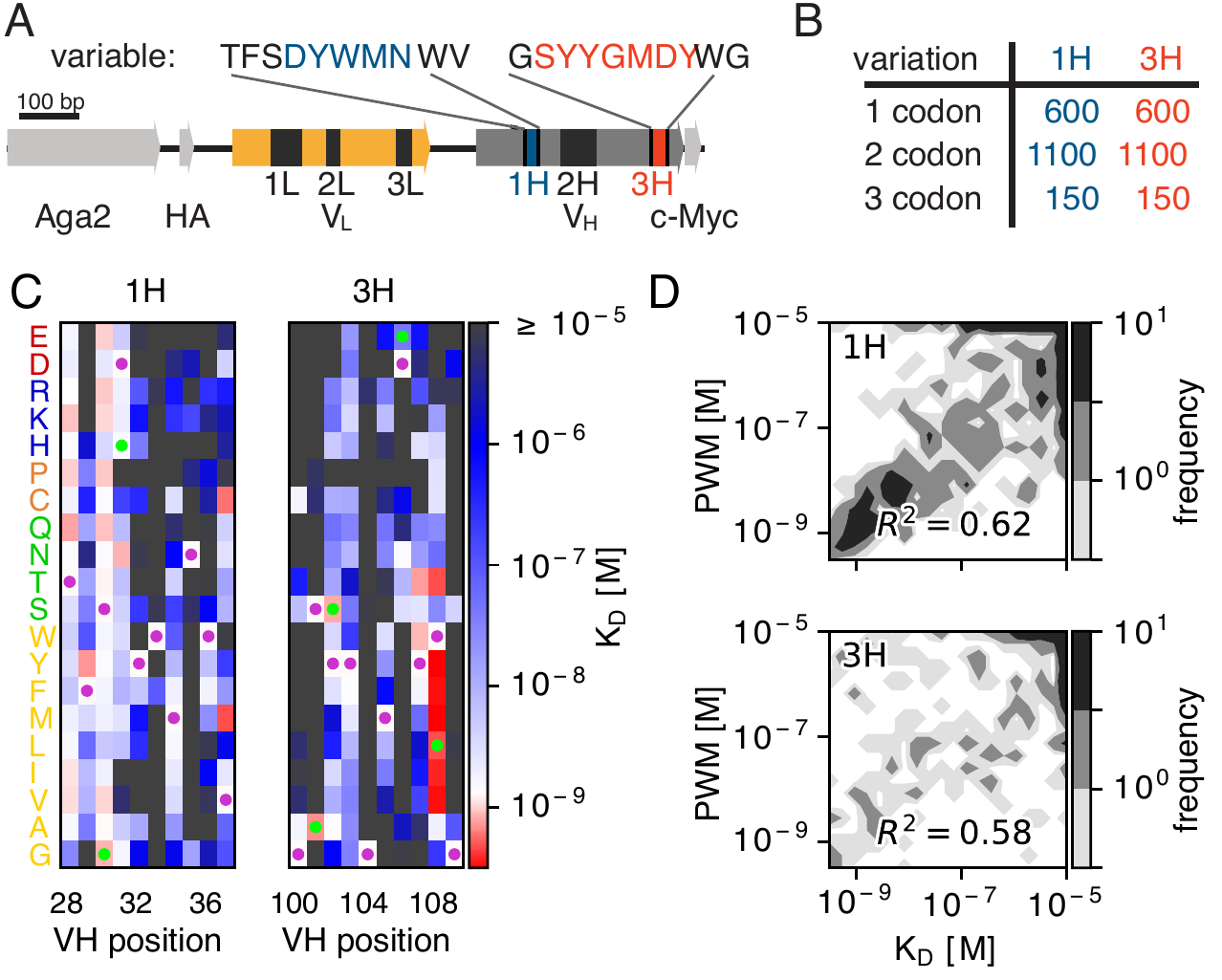}
\caption{{\bf Additive model of binding affinity.}
({\bf A}) 4-4-20 scFv antibody sequence. Six complementarity determining regions (CDR: 1L, 2L, 3L, 1H, 2H, 3H) are particularly important for antibody binding affinity.
({\bf B}) A library of antibody sequences with mutations in 10 amino-acid regions around the CDR1H and CDR3H domains were expressed using yeast display. Using Tite-Seq, the binding constants $K_D$ of all 600 single codon mutants, 1100 random double codon mutants, and 150 random triple codon mutants, were measured.
({\bf C}) 
The $K_D$ of single mutants for 1H and 3H domains were used to create position weight matrices (PWM) to predict the affinity of double and triple mutants.
({\bf D}) Comparison between the PWM prediction and the measurement of $K_D$ on double and triple mutants. The PWMs explained a significant portion of the variance, as quantified by the explained variance $R^2$ ($p<10^{-61}$ for CDR1H, $p<10^{-48}$ for CDR3H, F-test for reduction in variance due to PWM).
PWMs trained from the binding free energy, $F=\ln(K_D/c_0)$, outperformed PWM trained from $K_D$ (Fig.~\ref{fig:log_v_lin_PWM}). We looked for the optimal nonlinear transformation of $K_D$ maximizing the PWM fits (Methods and Fig.~\ref{fig:test_fit} for validation on simulated data) and found that PWMs perform almost ideally when trained from the binding free energy (see Fig.~\ref{fig:best_energy_fun_boundary}).
}\label{fig1}
\end{figure*}

\section*{Results}
\subsection*{Position Weight Matrix model of affinity}
We analyzed data from \cite{adams_measuring_2016} (\url{https://github.com/jbkinney/16_titeseq}), where Tite-Seq was applied to measure the binding affinities of variants of the 4-4-20 fluorescein-binding scFv antibody, henceforth called `wildtype'. Libraries were generated by introducing mutations to either the CDR1H or CDR3H domains restricted to 10 amino acid stretches called 1H and 3H (Fig.~\ref{fig1}A). All single amino acid mutants, 1100 random double amino acid mutants, and 150 triple amino acid mutants were generated in multiple synonymous variants and measured, (Fig.~\ref{fig1}B). Using a combination of yeast display and high-throughput sequencing at various antigen concentrations, Tite-Seq yielded the binding dissociation constant $K_D$ (in M or mol/L) of each variant with the fluorescein antigen.

We first tried to predict the $K_D$ of double and triple mutants from single mutant measurements. Mutations are expected to act on the binding free energy in an approximately additive way \cite{wells1990additivity,olson_comprehensive_2014}.
One may thus write the free energy of binding, $F=\ln (K_D/c_0)$ (defined up to constant in units of $k_B T$), as a sum over mutations in the mutagenized region, $\mathbf{s}=(s_1,\ldots,s_{\ell})$: 
\begin{align}\label{eq:PWM}
F(\mathbf{s})\approx F_{\rm PWM}(\bs) = F_{\rm{WT}} + \sum_{i=1}^\ell h_i(s_{i}),
\end{align}
where $F_{\rm WT}$ is the wildtype sequence energy, and $h_i(s_i)$ is the effect of a mutation at position $i$ to residue $s_i$.  The elements of the Position-Weight Matrix (PWM) $h_i(s)$ are obtained from the $K_D$ of single mutants shown in Fig.~\ref{fig1}C.
Since Tite-Seq measurements are limited to values of $K_D$ ranging from $10^{-9.5}$ to $10^{-5}$, for consistency PWM predictions outside this range were set to the boundary values.
The PWM was a fair predictor of double and triple mutants (Fig \ref{fig1}D), accounting for 62\% ($p<10^{-61}$, F-test) of the variance for 1H mutants and 58\% ($p<10^{-48}$, F-test) of the variance of 3H mutants.

The unexplained variance missed by the PWM model may have three origins: measurement noise, epistasis, or nonlinear effects. The last case corresponds to the hypothesis of additivity not being valid for $F=\ln(K_D/c_0)$, but for some other nonlinear transformation of $F$. Such a nonlinearity, also called ``scale,'' can lead to spurious epistasis \cite{Fisher1918,Phillips2008}. We first checked that additivity did not apply to the untransformed dissociation constant, $K_D$: a PWM model learned from $K_D$ instead of $F$ could only explain 34\% of the variance of all 1H and 3H multiple mutants, down from 62\% when learning from $F$ (Fig.~\ref{fig:log_v_lin_PWM}). We then looked for the non-linear transformation $E(F)$ that would give the PWM model with the best predictive power (Methods and Fig.~\ref{fig:test_fit}). This optimization yielded only a modest improvement to 65\% of the explained variance. In addition, the optimal function $E$ was very close to the logarithm ($R^2=97\%$, Fig.~\ref{fig:best_energy_fun_boundary}). Since nonlinear effects do not play a significant role, henceforth we only consider the PWM model defined on the free energy.

\begin{figure*}
\includegraphics[width=0.65\linewidth]{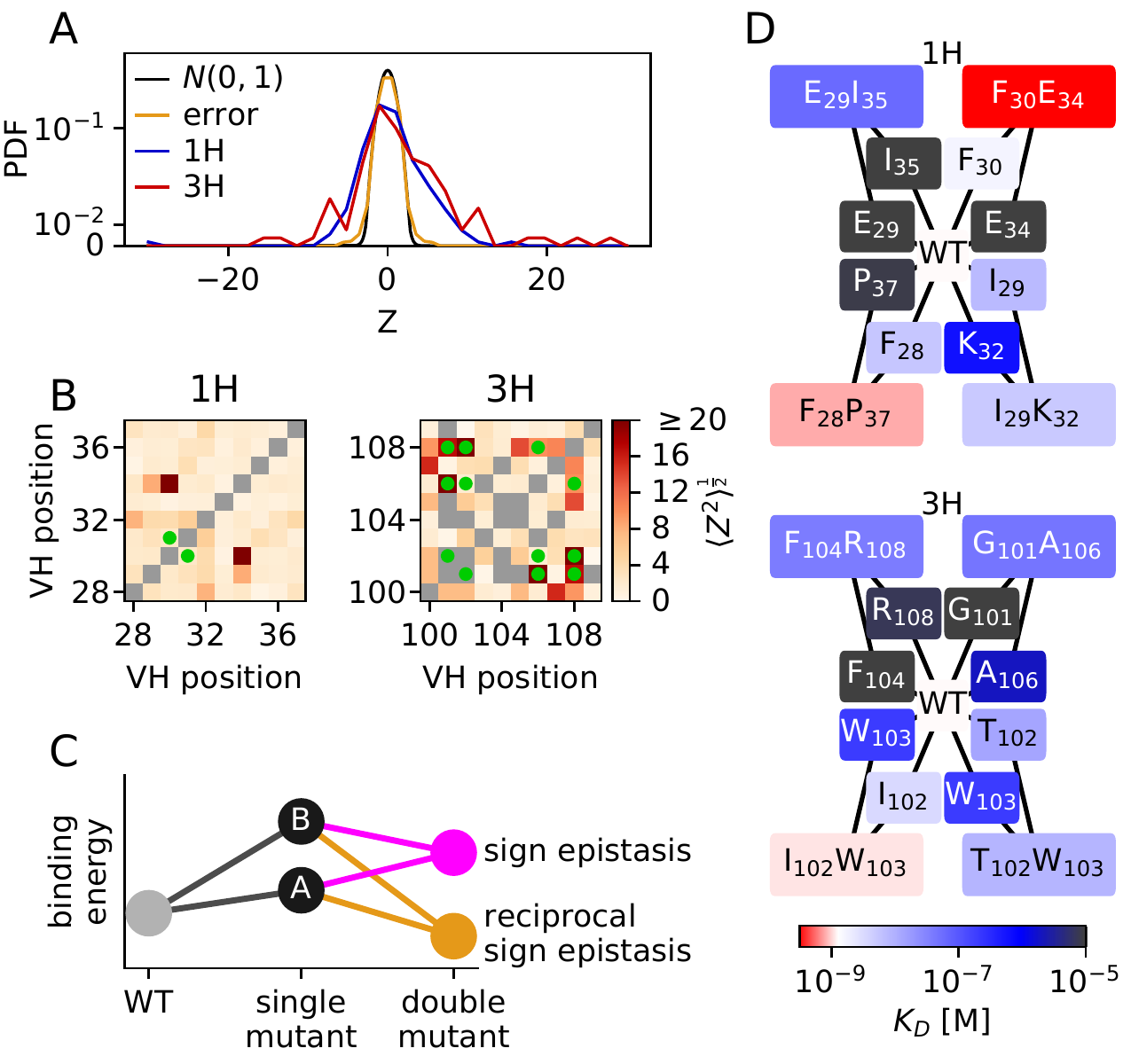}
\caption{{\bf Quantification of epistasis.}
Epistasis is defined as deviation from the PWM model, which assumes an additive effect of single mutations on the binding free energy $F= \ln (K_D/c_0)$ expressed in units of $k_BT$.
({\bf A}) Distribution of Z-scores, defined as the normalized deviation from the PWM prediction, $Z_{\rm epi}=({F_{\rm{PWM}}-F})/{\sqrt{\sigma^2 + \sigma_{\rm{PWM}}^2}}$, where $\sigma^2$ and $\sigma_{\rm{PWM}}^2$ are the estimated errors on $F$ and $F_{\rm PWM}$. Positive Z-scores indicate epistasis increased affinity.
The Z score standard deviation was much higher than expected from measurement errors ($Z_{\rm error}$)
for CDR1H (3.34, $p<10^{-33}$, Levene's test) and CDR3H (5.44, $p<10^{-52}$), meaning that the discrepancy between the PWM and measurement is mainly due to true epistasis. 
 ({\bf B}) Standard Z-score deviation for each pair of positions along the sequence. This deviation is higher at pairs of positions mutated in the super-optimized 4m5.3 antibody (green dots) in 3H ($p=0.005$, Mann-Whitney), but not in 1H ($p=0.23$).
Pairs of positions with large epistatic effects are shown on the wild-type crystal structure in Fig.~\ref{fig:epistasis_structure}. There is a weak correlation between epistasis and distance between the residues (Fig.~\ref{fig:contact_v_epistasis}).
({\bf C}) Schematic representation of sign and reciprocal sign epistasis for a beneficial interaction.
({\bf D}) Representative examples of sign epistasis as identified by Z-score. All \RA{44} examples of beneficial sign epistasis with double mutant $K_D\leq 10^{-6}$ may be found in \url{S1_table_sign_epistasis.csv}, and summarized in tables \ref{table:CDR1H_sign_epistasis} and \ref{table:CDR3H_sign_epistasis}.
Examples of deleterious sign epistasis are shown in figure \ref{fig:deleterious_sign_epistasis}.
}\label{fig2}
\end{figure*}

\subsection*{Epistasis affects affinity \label{section:model_contributions}}
To identify epistasis, we estimated the difference between the measured binding free energies of double and triple mutants, $F(\bs)$, and the PWM prediction, $F_{\rm PWM}(\bs)$. However, these small differences can be confounded by measurement noise, which can be mistaken for epistasis. To control for this noise, we defined Z-scores between two estimates of the free energy, $F_a$ and $F_b$, as
$Z = ({F_a - F_b})/{\sqrt{\sigma_a^2 + \sigma_b^2}}$,
where $\sigma_a^2$ and $\sigma_b^2$ are their estimates of uncertainty. 
We first computed Z-scores between independent estimates of the same free energy using synonymous variants ($Z_{\rm error}$, Methods). We found that the distribution of $Z_{\rm error}$ was normal with variance $\approx 1$ (Fig.~\ref{fig2}A, orange line), as expected from Gaussian measurement noise.

We then estimated the effect of epistasis by calculating Z-scores ($Z_{\rm epi}$) from the difference between the PWM prediction, $F_{\rm PWM}$ (Eq.~\ref{eq:PWM}), and the measured $F$. The resulting distributions of Z-scores (Fig.~\ref{fig2}A, blue and red lines) had much larger variances than expected from measurement noise (standard deviation 3.34 for 1H, and 5.44 for 3H), indicating strong epistasis. These epistatic effects were on average slightly beneficial (positive $Z$):  {25}\% of double mutants inside the reliable readout boundaries ($10^{-9.5}M \leq K_D\leq10^{-5} M$) showed significant beneficial epistasis ($Z_{\rm epi}>1.64$, $p=0.05$), and {20}\% significant deleterious epistasis ($Z_{\rm epi}<-1.64$).
Comparing the variance of $Z_{\rm epi}$ with that of $Z_{\rm error}$ gives a large fraction of the unexplained variance that is attributable to epistasis, $1-{\mathrm{Var}(Z_{\rm error})}/{\mathrm{Var}(Z_{\rm epi})}= 89\%$ for 1H, and 96\% for 3H.

To determine whether certain positions along the sequence concentrated epistatic effects, we computed the mean squared Z-score for all double mutations at each pair of positions (excluding median boundary values), revealing a complex and heterogeneous landscape of epistasis (Fig.~\ref{fig2}B and Fig.~\ref{fig:epistasis_structure} for the epistasis magnitude superimposed on the wildtype's crystal structure). CDR3H, which interacts directly with the antigen, is observed to have more epistatically interacting sites than CDR1H. Interestingly, the three most epistatic pairs in 3H --- between positions 101, 106 and 108 --- are mutated in the previously described super-optimized 4m5.3 antibody \cite{boder_directed_2000} (mutations shown in green in Fig.~\ref{fig1}B), consistent with previous suggestions that positions 101 and 106 interact together and with position 108 via hydrogen bonds \cite{Midelfort:2004iq,adams_measuring_2016}.
Epistasis is usually expected between residues that are in contact in the protein structure \cite{romero_navigating_2013,morcos_direct-coupling_2011,mclaughlin_spatial_2012,zhang_evolution_2013,melamed_deep_2013}, as for instance between positions 101 and 106. However, the mean squared Z-score weakly correlated with residue distance ($r=-0.13,p=0.21$ for 1H, $r=-0.34,p=0.003$ for 3H, Fig.~\ref{fig:contact_v_epistasis}).

We next looked for evidence of ``sign epistasis,'' where one mutation reverses the sign of the effect of another mutation (Fig.~\ref{fig2}C). We defined a Z-score for a single mutation A quantifying the beneficial effect of that mutation relative to the noise,
$Z_{A} = ({F_{\rm{WT}} - F_A})/{\sigma_A}$,
where $F_{\rm{WT}}$ and $F_A$ are the wildtype and mutant free energies, and $\sigma$ is the measurement error estimated as before. Since we are only interested in the sign of the effect, we keep single mutants at the reliable readout boundary.
An equivalent Z-score was defined for a mutation A in the background of an existing mutation B: $Z_{A|B}=(F_B-F_{AB})/\sqrt{\sigma_A^2+\sigma_{AB}^2}$, where $F_{AB}$ is the free energy of the double mutant AB. Significant sign epistasis was defined by $Z_{A|B}Z_A<0$ and $|Z_{A|B}|,|Z_{A}|>1.64$, and reciprocal sign epistasis by the additional symmetric condition $A\leftrightarrow B$.

We found 44 cases of significant epistasis, listed in \url{S1_table_sign_epistasis.csv} and summarized in Tables \ref{table:CDR1H_sign_epistasis} and \ref{table:CDR3H_sign_epistasis}. Deleterious sign epistasis was exceptional, with just one instance in 1H and 4 in 3H (Fig.~\ref{fig:deleterious_sign_epistasis}). The four most significant cases of beneficial sign epistatis for each domain are depicted in Fig.~\ref{fig2}D.
Among cases where both single mutations were deleterious, we found $3\%$ of mutants in 1H and $0.7\%$ in 3H with significant beneficial epistasis, versus $0.06\%$ expected by chance (the null expectation, which takes into account the constraint that $Z_A+Z_{B|A}=Z_B+Z_{A|B}$, is defined in the Methods); $0.7\%$ were reciprocal in 1H, and $0.3\%$ in 3H, versus $0.01\%$ expected by chance. To evaluate how these epistatic interactions may affect affinity maturation, we estimated how often ``viable'' double mutants were separated from the wildtype by nonviable single mutants, where viability is defined by $K_D<10^{-6}$M \cite{batista1998,foote1995kinetic,roost1995early}, forming possible roadblocks to affinity maturation. This strong instance of ``rescue'' epistasis occured in roughly half of the mutants with beneficial sign epistasis (Table \ref{table:CDR1H_sign_epistasis} and \ref{table:CDR3H_sign_epistasis}).

\begin{figure*}
\includegraphics[width=0.65\linewidth]{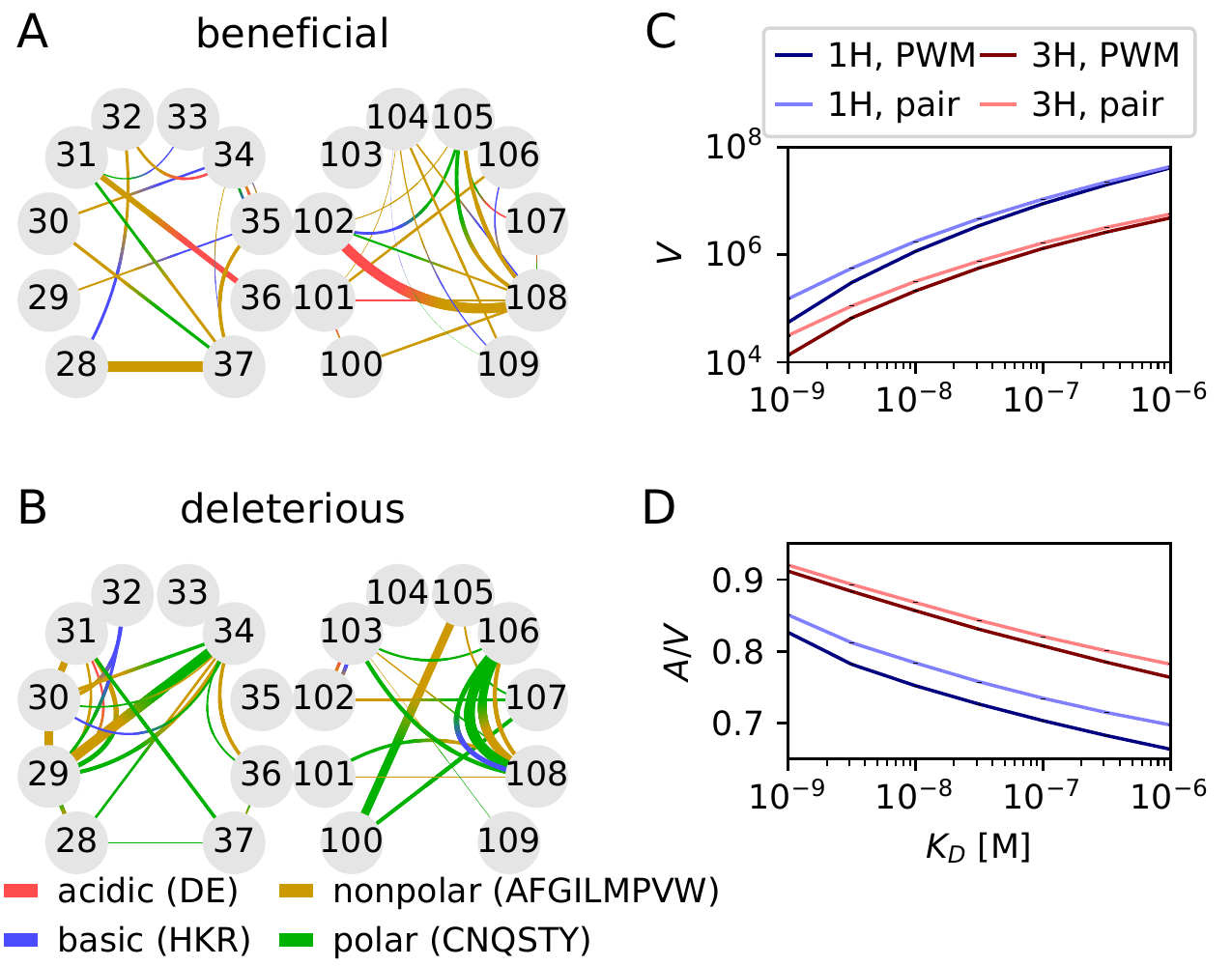}
\caption{{\bf Coarse-grained epistatic model.} A model of biochemical epistatic interactions between polar, nonpolar, acidic, and basic residues was fitted to the data using LASSO regularization and tested by cross-validation (Fig.~\ref{fig:bayesian_p_vals}A), yielding 1058 CDR1H and 1066 CDR3H interaction terms. Mean ({\bf A}) beneficial and ({\bf B}) deleterious interactions calculated by averaging over all double mutants, colored by interaction type. Line width denotes interaction strength. The model performance on significantly epistatic pairs of positions is shown in Figs.~\ref{fig:bayesian_p_vals}B-C, and the number of non vanishing parameters as a function of the significance threshold on the posterior is shown in Fig.~\ref{fig:bayesian_p_vals}D.
({\bf C}) Number $V$ of amino-acid sequences of the 1H (blue) and 3H (red) regions with dissociation constant below $K_D$, as estimated by the PWM model (dark color) or the epistatic model (light color). Epistasis enlarges the number of variants with good affinity for both 1H and 3H.
({\bf D}) Mutational flux $A$ (defined as the average number of random mutation events from all possible sequences to cause the dissociation constant to cross $K_D$), normalized by $V$, showing that epistasis also increases the accessibility of the region of good binders in sequence space. Differences between the PWM and epistatic models were robust to errors in the estimate of the interaction parameters ($p<10^{-5}$, Jackknife analysis).
}\label{fig3}
\end{figure*}

\subsection*{Modeling epistasis and its impact on affinity maturation}
To integrate the observed epistatic interactions into a predictive model of affinity, we introduced a model of binding free energy as:
\begin{align}
F(\bs)\approx F_{\rm pairwise}(\bs) = F_{\rm PWM}(\bs) + \sum_{i<j} J_{ij}(s_i,s_j)\label{eq:pairwise},
\end{align}
where $J$ is the interaction strength between residues. To avoid overfitting and allow for independent validation (in the absence of a sufficient number of triple mutants), we grouped residues into 4 biochemical categories \cite{voet2011biochemistry} (polar, nonpolar, acidic, basic, see Methods) and let the entries of $J$ only depend on that category.

We trained the model on the 1208 1H or 1216 3H double and triple mutants, using a Lasso penalty to control for overfitting. The optimal penalty was set by 10 fold cross-validation, i.e. by maximizing the explained variance of a subset comprising $1/10$ of the mutants by using a model trained on the remaining $9/10$, averaged over the 10 subsets (Fig.~\ref{fig:bayesian_p_vals}A and Methods). Interacting pairs with posterior probabilities $>0.95$ as determined by Bayesian Lasso \cite{park_bayesian_2008} are shown in Figs.~\ref{fig3}A and B.

Out of the 720 possible terms, 52 1H and 45 3H interaction terms were identified by this method. Although these interactions, whose number is limited by the number of measured variants, only modestly improved the explained variance relative to the PWM in all multiple mutants (from {62}\% to {64}\% for 1H and from {58}\% to {60}\% for 3H), it substantially improved the affinity prediction of the mutants with significant epistasis ($R^2$ from 27\% to 50\% in 1H, from 13\% to 44\% for 3H, Fig.~\ref{fig:bayesian_p_vals}B-C).
Notably, two mutations of the super-optimized 4m5.3 antibody are predicted by the model to have epistatic interactions: a slightly deleterious effect between $A_{101}$ and $L_{108}$, and a strongly beneficial one between $S_{102}$ and $L_{108}$.

Next we used our models to estimate the diversity, or ``degeneracy'', of antibodies with good binding affinity. Specifically, we evaluated the degeneracy volume $V$ of high-affinity sequences as the number of sequences with $K_D<B$, using either the PWM (Eq.~\ref{eq:PWM}) or pairwise (Eq.~\ref{eq:pairwise}) models, using a combination of exhaustive and Monte-Carlo sampling (Methods).
Compared to the coarse-grained pairwise model trained previously, the interaction strength $J$ was learned directly for each residue pair, without grouping by biochemical category and with no Lasso penalty. The volume of 1H mutants was larger than that of 3H mutants (Fig.~\ref{fig3}C), in agreement with the fact that CDR3H plays a more important role in binding affinity. Epistasis increased the recognition volume for both domains, consistent with the previous observation that epistatic effects are, on average, more beneficial than deleterious.
To explore the diversity of evolutionary paths leading to recognition, we computed the mutational flux $A$ in and out of the high-affinity region as the probability that a random mutation in a high-affinity sequence ($K_D<B$) causes loss of recognition ($K_D>B$), summed over all high-affinity sequences (Methods). Again we found that epistasis increased the mutational flux, even after normalizing by volume, $A/V$ (Fig.~\ref{fig3}D).
We checked that these differences were robust to sampling noise and overfitting by performing a jackknife analysis ($p<10^{-5}$ for the difference in $A$ and $V$ between the PWM and pairwise models, see Methods).

 % discussion
\section*{Discussion}
By analyzing massively parallel affinity measurements obtained by Tite-Seq, we painted a detailed picture of epistasis in a well-defined physical phenotype --- the binding free energy of an antibody to an antigen. We showed that antibody sequences contain many epistatic interactions contributing to the binding energy, and that many of them have beneficial effects.  Our approach involves first training an additive (PWM) model as a baseline, and identifying departures from that model as epistasis. In this comparison, a crucial step was to correct for the two issues of scale and measurement noise.

The first issue, identified by Fisher \cite{Fisher1918} and also called unidimensional epistasis \cite{Szendro2013}, is the idea that an epistatic trait becomes additive upon a different parametrization. For instance, protein stability, which often determines fitness, is a nonlinear function of the folding free energy difference, which is expected to be roughly additive \cite{Bloom2005,Bershtein2006,Jacquier2013,Gong2013,Serohijos2014,Bank2015,Sarkisyan2016}. This leads to both a law of diminishing returns \cite{Bank2015} and robustness to mutations when the protein is very stable \cite{Bloom2005}. To disentangle these potential artifacts, we defined our PWM on the binding free energy, which we expect to be additive in sequence content, and we checked that this parametrization was close to minimizing epistasis.

To tackle the second and perhaps more important issue of noise, especially in the context of deep mutational scans where many variants are tested \cite{Araya:2012p13011}, we developed a robust methodology based on Z-scores to identify epistatic interactions as significant outliers. This analysis showed that almost all of the variance unexplained by additivity ($\sim 40\%$) could be attributed to epistasis, making its contribution to the phenotype comparable to that of single mutations. A large fraction of that epistasis was beneficial, in contrast with previous reports of mostly negative epistasis owing to the concavity of the scale \cite{Bershtein2006,Schenk2013,Bank2015}, which we here circumvent by directly considering the free energy.

Epistasis is key to understanding the predictability and reproducibility of evolutionary paths \cite{lassig_predicting_2017,kryazhimskiy_global_2014}. Our results show how it could constrain the space of possible hypermutation trajectories during affinity maturation, with important consequences for antibody and vaccine design, as the importance of eliciting responses of antibodies that are not just strongly binding but also evolvable is being increasingly recognized \cite{Wang2015a}.
Targeting epistatic interactions may provide an alternative strategy for optimizing antibody affinity: among the 12 epistatic hotspots in CDR1H and 20 in CDR3H that we identified ($\langle Z_{\rm{epi}}^2\rangle^{\frac{1}{2}}>3$), 4 involved positions mutated in the super-optimized 4m5.3 antibody sequence, with a higher epistatic contribution than expected by chance. We also identified 3 cases of beneficial sign epistasis, in which the double mutant was fit despite the single mutant being deleterious. For instance, the D108E mutations in 4m5.3 is deleterious by itself but is rescued beyond the wildtype value by the S101A mutation \cite{Midelfort:2004iq}, which occurred first in the directed evolution process \cite{boder_directed_2000}. We report 10 extreme cases of viable double mutants whose single-mutant intermediates are nonviable, possibly blocking affinity maturation.
However, our analysis of the volume and mutational flux of the region of low binding free energies in sequence space suggests that epistasis facilitates the evolution of high-affinity antibodies. Additionally interactions with the non-mutated parts of the sequence and evolution of the antigen binding partner can either add further constraints or open up additional paths. 

Taken together, our results show the importance of taking into account epistasis when predicting antibody evolution and guiding vaccine design.  Our systematic approach for identifying and quantifying epistasis, with the implementation of important controls for scale and noise, could be used by other investigators to analyze deep-mutational scans of protein function.

 % methods
\section*{Methods}
Values of $K_D$ as measured by Tite-Seq for variants of the 4-4-20 fluorescein-binding antibody \cite{adams_measuring_2016} can be found at \url{https://github.com/jbkinney/16_titeseq}. The scripts used for the analyses presented here are available at \url{https://github.com/rhys-m-adams/epistasis_4_4_20}.
\subsection*{Position Weight Matrix}
The amino-acid sequence of the 10 amino acid stretches of the CDR1H or CDR3H domains are denoted by $\bs=(s_1,\ldots,s_{10})$. The corresponding 30-long nucleotide sequences are denoted by $\bv$. The binding free energy $F(\bs)$ of an amino-acid variant is obtained as the mean over 3 replicate experiments, and over all its synonymous variants:
\begin{equation}\label{defF}
F(\bs) =  \frac{1}{N(\bs)}\sum_a  \sum_{\bv \in S_a(\bs)} \ln(K_D(\bv,a)/c_0),
\end{equation}
where $S_a(\bs)$ is the set of measured nucleotide sequences that translate to $\bs$ in replicate $a$, and $N(\bs)=\sum_a |S_a(\bs)|$ a normalization constant.

The elements of the PWM are defined as $h_i(q)=F(\bs^{(i,q)})-F_{WT}$, where $\bs^{(i,q)}$ is the single mutant mutated at position $i$ to residue $q$, and $h_i(q)=0$ when $q$ is the wildtype residue at position $i$.

\subsection*{Optimal nonlinear transformation of the free energy \label{section:best_energy_fun}}
To test whether transforming $F$ through a nonlinear function $E(F)$ before learning the PWM could improve its predictive power, we defined the nonlinear additive model:
\begin{equation}
F(\bs)\approx f[E_{\rm PWM}(\bs)],\quad E_{\rm PWM}(\bs)=E_{\rm WT}+\sum_i \tilde h_i(s_i),
\end{equation}
where $f=E^{-1}$ is the inverse function of $E$, $\tilde h_i(q)=E(\bs^{(i,q)})-E_{\rm WT}$, and $E(\bs)$ is evaluated similarly to Eq.~\ref{defF}: $E(\bs)=(1/N(\bs))\sum_a \sum_{\bv \in S_a(\bs)} E[\ln(K_D(\bv)/c_0)]$.

To find the transformation $E$ that gives the highest explained variance while avoiding overfitting, we aimed to minimize the following objective function:
\begin{align}\label{objective}
O[E]=\sum_{\bs} \left[ E_{\rm{PWM}}(\bs)-E(\bs)\right ]^2 + \alpha \int dF\, |E''(F)|^2,
\end{align}
where the sum in $\bs$ runs over double and triple mutants, and $\alpha$ is a tunable parameter.

Numerically, we parametrize the function $E(F)$ as piecewise linear: $E(F)=E_i\times (F_{i+1}-F)/\delta F + E_{i+1}\times (F-F_{i})/\delta F$ for $F_i\leq F\leq F_i$, where $F_i$ are equally spaced grid point along $F$, $\delta F=F_{i+1}-F_i$, and $E_i$ the value of $E$ at these points. The smoothing penalty is approximated by a sum over the squared discretized second derivative: $\int dF |E''(F)|^2\approx \sum_i (E_{i+1}+E_{i-1}-2E_i)^2/\delta F^3$.

We minimize $O[E]\approx O[E_1,\ldots,E_N]$ as a quadratic function of its arguments $(E_i)$, while imposing boundary constraints on the PWM prediction and the requirement that $E$ is a increasing function of $F$ (i.e. $E_{i+1}>E_i$), using the python package {\tt cvxopt} \cite{andersen2013cvxopt}.
 
The hyper-parameter $\alpha$ is evaluated by maximizing the generalized cross-validation of the coefficient of determination 
\begin{equation}
R^2= 1-\left\langle\frac{\sum_{\bs\in S} [ E^{\backslash S}_{\rm PWM}(\bs)-E^{\backslash S}(\bs) ]^2}{\mathrm{Var}_{\bs\in S} [E^{\backslash S}(\bs)]}\right\rangle_S,
\end{equation}
where $E^{\backslash S}$ and $E^{\backslash S}_{\rm PWM}(\bs)$ are learned through optimizing Eq.~\ref{objective}, but after removing from the dataset a subset $S$ of the multiple mutants comprising one tenth of the total. The average is over ten non-overlapping subsets $S$.

This method was first tested on simulated data. Each PWM element $\tilde h_i(q)$ was drawn from a normal distribution of zero mean and variance 1, and then $E_{\rm PWM}(\bs)$ was computed for each of the antibody sequences present in our data. Our simulated ``measurement'' was defined as a function of a noisy realization of $E=E_{\rm PWM}+\epsilon$ (where $\epsilon$ is some Gaussian noise)
in three different ways: linear $F=E$, exponential $F=\exp(E)$, high-frequency $F = 2E+ \sin(2E)$, and logistic $F=1/[1+\exp(-E)]$.
$\epsilon$ was drawn from a centered normal distribution with 1/2 the standard deviation of $E_{\rm PWM}$. $F$ was then truncated to the 200th lowest and 200th highest values, to mimick the boundary cutoff in our measurements. Comparing our original $E_{\text{PWM}}$ to our fit $\hat{E}$ shows that our method is able to infer the true PWM model and a smooth nonlinearity from noisy data (Fig.~\ref{fig:test_fit}). 

We then applied the method to the experimental data. The cross-validation $R^2$ is represented as a function of the smoothing parameter $\alpha$ in Fig.~\ref{fig:best_energy_fun_boundary}A, and the corresponding optimal function $E(F)$ in Fig.~\ref{fig:best_energy_fun_boundary}B. The comparison between measurement and the PWM model is shown in Fig.~\ref{fig:best_energy_fun_boundary}C.

\subsection*{Z-scores}
We used synonymous mutants to estimate our measurement error. The mean free energy of a nucleotide sequence is defined as the mean over replicate measurements: $F(\bv)=\langle\ln(K_D(\bv,a)\rangle_a$, and the standard error $\sigma(\bv)$ is defined accordingly as the pooled error over replicates.
Antibodies with $K_D$ having median values at the boundary values of $10^{-9.5}$ or $10^{-5}$ were excluded from the analysis since these values artificially cluster at the boundary, leading to underestimates of error.

The error Z-score was calculated between pairs of nucleotide sequences with the same amino acid translation:
$Z_{\text{err}} (\bv, \bv')= ({F(\bv)-F(\bv')})/{\sqrt{\sigma(\bv)^2 + \sigma(\bv')^2}}$.

Epistatic Z-scores were estimated by calculating the measurement error over both replicates and synonymous variants, as in Eq.~\ref{defF}:
\begin{equation}
\sigma^2(\bs) =\frac{\sum_a  \sum_{\bv \in S_a(\bs)} \left[\ln(K_D(\bv,a)/c_0)-F(\bs)\right]^2}{N(\bs)(N(\bs)-1)},
\label{eqn:standard_error_definition}
\end{equation}
and the pooled standard error for a PWM prediction, calculated as the sum of measurement errors from single mutations:
\begin{equation}
\sigma_{\rm{PWM}}^2(\bs) = \sum_i \sigma_i^2(s_i),
\label{eqn:PWM_error_model}
\end{equation}
where $\sigma_i(q)=\sigma(\bs^{(i,q)})$, and $\sigma_i(q)=0$ when $q$ is the wildtype residue at $i$. The epistatic Z-score is defined as:
\begin{equation}
Z_{\rm epi}(\bs)=\frac{F_{\rm PWM}(\bs)-F(\bs)}{\sqrt{\sigma^2(\bs)+\sigma_{\rm PWM}^2(\bs)}}.
\end{equation}

\subsection*{Null model for sign epistasis}
To calculate p-values for sign epistasis, we used the following null model for sets of four Z-scores satisfying $Z_A+Z_{B|A}=Z_B+Z_{A|B}$. Calling $x_1=Z_A,x_2=Z_{B|A},x_3=-Z_{A|B},x_4=-Z_{A}$, the condition becomes that each $x_i$ has zero mean and variance one, with the constraint $\sum_{i=1}^4x_i=0$. The distribution with maximum entropy satisfying these requirements is a centered multi-variate Gaussian uniquely defined by its covariance matrix $\langle x_i^2\rangle=1$ and $\langle x_ix_j\rangle=-1/3$ for $i\neq j$. The p-value for sign epistasis, $Z_A>1.64$ and $Z_{A|B}<1.64$, was estimated by Monte Carlo sampling under that Gaussian distribution as $\mathrm{Pr}(x_1>1.64\ \&\ x_2>1.64)+\mathrm{Pr}(x_3>1.64\ \&\ x_4>1.64) - \mathrm{Pr}(x_1>1.64 \ \&\  x_2<-1.64\ \&\ x_3>1.64\ \&\ x_4<-1.64) = 6.2 \cdot 10^{-4}$, and the probability for reciprocal sign epistasis as $\mathrm{Pr}(x_1>1.64 \ \&\  x_2<-1.64\ \&\ x_3>1.64\ \&\ x_4<-1.64)=10^{-4}$.

\subsection*{Epistatic model \label{section:epistasis_model}}
The epistatic terms of the pairwise model were made to depend on the biochemical categories of the interacting residues, $J_{ij}(s_i,s_j)=\tilde J_{ij}(b(s_i),b(s_j))$, with $b(s)={\rm nonpolar}$ for $s={\rm AFGILMPVW}$, $b(s)={\rm polar}$ for $s={\rm CNQSTY}$, $b(s)={\rm acidic}$ for $s={\rm DE}$, and $b(s)={\rm basic}$ for $s={\rm HKR}$. A fifth category was added to correspond to the wildtype residue, so that $\tilde J_{ij}({\rm wildtype},b)=\tilde J_{ij}(b,{\rm wildtype})=0$.
The model was trained by minimizing the mean squared error with a regularization penalty over all matrices $\tilde J_{ij}(b,b')$:
\begin{align}
\sum_{\bs} 
\left [F(\bs) - {F}_{\rm{pairwise}}(\bs, {\tilde J})\right ]^2
+ \lambda \sum_{ijbb'}|\tilde J_{ij}(b,b')|.
\end{align}
The Lasso penalty $\lambda$ was learned by 10-fold cross-validation, and energy terms found in less than 2 sequences were excluded from the fit. Posterior values for $\tilde J$ terms were calculated using Bayesian Lasso \cite{park_bayesian_2008}.

The volume and mutational flux were defined as:
\begin{align}
V(B)&=\sum_{\bs}\Theta(B-K_D(\bs))\label{VB}\\
A(B)&=\sum_{\bs}\Theta(B-K_D(\bs)) \frac{1}{19\ell}\sum_{\bs'|d(\bs,\bs')=1}\Theta(K_D(\bs')-B),\label{AB}
\end{align}
where $\Theta(x)$ is the Heaviside function, i.e. $\Theta(x)=1$ if $x\geq 0$ and $0$ otherwise; $d(\bs,\bs')$ is the Hamming distance between two sequences; and $\ell=10$ is the sequence length. The normalization $19\times \ell$ corresponds to the number of mutants $\bs'$ at Hamming distance 1 from $\bs$. The sums over $\bs$ in Eqs.~\ref{VB}-\ref{AB} have $20^{10}$ elements and are computationally intractable. To overcome this, we approximated the sum using a mixture of Monte-Carlo and complete enumeration, depending on the distance of $\bs$ from the wildtype. Calling $C_d$ the set of sequences $\bs$ at Hamming distance $d$ from wildtype, we used:
\begin{equation}
\sum_\bs g(\bs) \approx \sum_{d=0}^\ell \frac{|C_d|}{|\tilde C_d|}\sum_{\bs \in \tilde C_d} g(\bs),
\end{equation}
where $g(\bs)$ is a function of $\bs$ to be summed such as in $V$ or $A$ in Eqs.~\ref{VB}-\ref{AB}, and $\tilde C_d$ is a random subset of $C_d$ of size $\min(|C_d|,P_d)$, with $P_d=\binom{\ell}{d} \times (\floor{P/\binom{\ell}{d}}+1) $, where $P$ is the maximum number of sequences one is willing to sample completely at each $d$ to perform the estimation, and where $|C_d|=\binom{\ell}{d}19^d$. For small $d$, when $|C_d|\leq P_d$, the enumeration is complete, while for large $d$ and $|C_d|>P_d$, the sum is estimated from a uniformly distributed Monte Carlo sample of $C_d$.

\acknowledgments{}
We would like to thank Yuanzhe Guan and Carlos Talaveira for their suggestions. The authors declare no conflicts of interest. R.M.A., T.M. and A.M.W. were supported by grant ERCStG n. 306312. 
\bibliographystyle{pnas}

\clearpage

\beginsupplement

 % si
\begin{table*}
 % CDR1H_sign_epistasis.tex
\begin{center}
\begin{tabular}{ | l | l | l | l | l | l | l | l | l | l |}
\hline

domain &
\multicolumn{1}{|p{2cm}|}{\centering \# of nonviable single mutations} &
\multicolumn{1}{|p{2cm}|}{\centering epistasis type } &
\multicolumn{1}{|p{2cm}|}{\centering \# candidate mutants } &
\multicolumn{1}{|p{2cm}|}{\centering \# of mutants with sign epistasis (obs/exp)} &
\multicolumn{1}{|p{2cm}|}{\centering \# of mutants with reciprocal sign epistasis (obs/exp)} &
\multicolumn{1}{|p{2cm}|}{\centering \# of viable mutants with sign epistasis (obs/exp)} &
\multicolumn{1}{|p{2cm}|}{\centering \# of viable mutants with reciprocal sign epistasis (obs/exp)} &
\multicolumn{1}{|p{2cm}|}{\centering \# of sign epistatic mutants with $K_D <$ WT (obs/exp)}
  \\ \hline 
1H & $0$ & beneficial & 293 & 2/0.21  & 0/0.04 & 2/0.06 & 0/0.01 & 0/0.01 \\ \hline 
1H & $1$ & beneficial & 532 & 16/0.32  & 0/0.05 & 16/0.09 & 0/0.02 & 2/0.01 \\ \hline 
1H & $0, 1$ & beneficial & 825 & 18/0.53  & 0/0.09 & 18/0.15 & 0/0.03 & 2/0.02 \\ \hline 
1H & $2$ & beneficial & 172 & 14/0.10  & 7/0.02 & 12/0.03 & 6/0.01 & 0/0.01 \\ \hline 
1H & $0-2$ & beneficial & 997 & 32/0.63  & 7/0.11 & 30/0.18 & 6/0.03 & 2/0.02 \\ \hline 
1H & $0-2$ & deleterious & 53 & 1/0.63  & 0/0.11 & 0/0.18 & 0/0.03 & 0/0.02 \\ \hline 
    \end{tabular} 

\end{center}

\caption[Table caption text]{Summary of CDR1H sign epistasis with double mutants. Obs/exp denotes observed sign epistasis events versus expected epistasis events in a Gaussian noise model. Expected numbers below 0.01 are rounded up to 0.01. We measured 1058 CDR1H double mutants in total.}
\label{table:CDR1H_sign_epistasis}
\end{table*}

\begin{table*}
 % CDR3H_sign_epistasis.tex
\begin{center}
\begin{tabular}{ | l | l | l | l | l | l | l | l | l | l |}
\hline

domain &
\multicolumn{1}{|p{2cm}|}{\centering \# of nonviable single mutations} &
\multicolumn{1}{|p{2cm}|}{\centering epistasis type } &
\multicolumn{1}{|p{2cm}|}{\centering \# candidate mutants } &
\multicolumn{1}{|p{2cm}|}{\centering \# of mutants with sign epistasis (obs/exp)} &
\multicolumn{1}{|p{2cm}|}{\centering \# of mutants with reciprocal sign epistasis (obs/exp)} &
\multicolumn{1}{|p{2cm}|}{\centering \# of viable mutants with sign epistasis (obs/exp)} &
\multicolumn{1}{|p{2cm}|}{\centering \# of viable mutants with reciprocal sign epistasis (obs/exp)} &
\multicolumn{1}{|p{2cm}|}{\centering \# of sign epistatic mutants with $K_D <$ WT (obs/exp)}
  \\ \hline 
3H & $0$ & beneficial & 173 & 2/0.11  & 0/0.02 & 2/0.01 & 0/0.01 & 1/0.01 \\ \hline 
3H & $1$ & beneficial & 525 & 2/0.31  & 0/0.05 & 0/0.04 & 0/0.01 & 0/0.01 \\ \hline 
3H & $0, 1$ & beneficial & 698 & 4/0.42  & 0/0.07 & 2/0.05 & 0/0.01 & 1/0.01 \\ \hline 
3H & $2$ & beneficial & 359 & 3/0.21  & 3/0.04 & 3/0.03 & 3/0.01 & 0/0.01 \\ \hline 
3H & $0-2$ & beneficial & 1057 & 7/0.63  & 3/0.11 & 5/0.08 & 3/0.01 & 1/0.01 \\ \hline 
3H & $0-2$ & deleterious & 97 & 4/0.63  & 0/0.11 & 3/0.08 & 0/0.01 & 0/0.01 \\ \hline 
    \end{tabular} 

\end{center}

\caption[Table caption text]{Summary of CDR3H sign epistasis with double mutants. Obs/exp denotes observed sign epistasis events versus expected epistasis events in a Gaussian noise model. Expected numbers below 0.01 are rounded up to 0.01. We measured 1066 CDR3H double mutants in total. }
\label{table:CDR3H_sign_epistasis}
\end{table*}

\begin{figure*}
\includegraphics[width=0.65\linewidth]{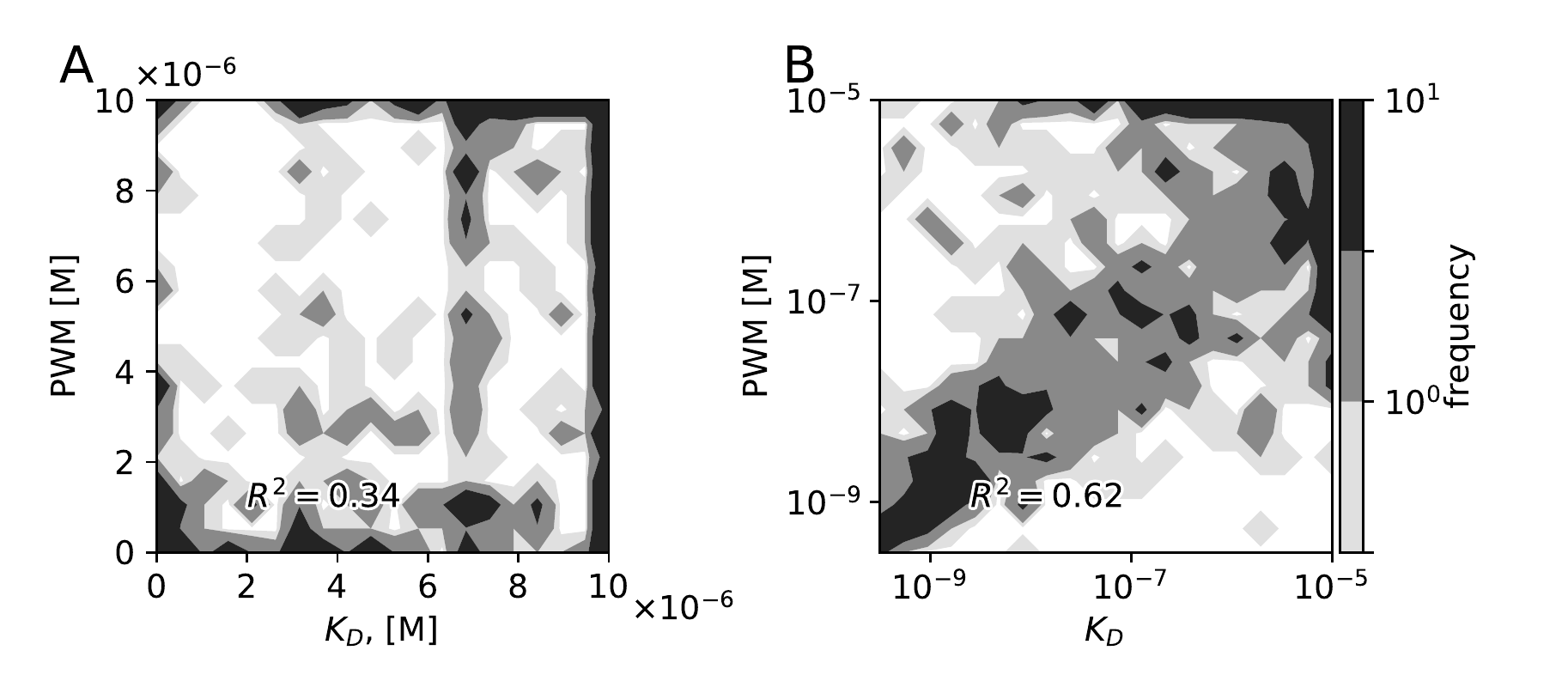}
\caption{Comparison between data and model prediction of the binding affinity of multiple mutants using PWMs constructed from A) $K_D$ and B) $F=\ln(K_D/c_0)$. $R^2$ denotes the coefficient determination (fraction of explained variance). Although both predictions are very significant ($p<10^{-5}$, F-test), the PWM based on $F$ is much better.}\label{fig:log_v_lin_PWM}
\end{figure*}

\begin{figure*}
\includegraphics[width=0.65\linewidth]{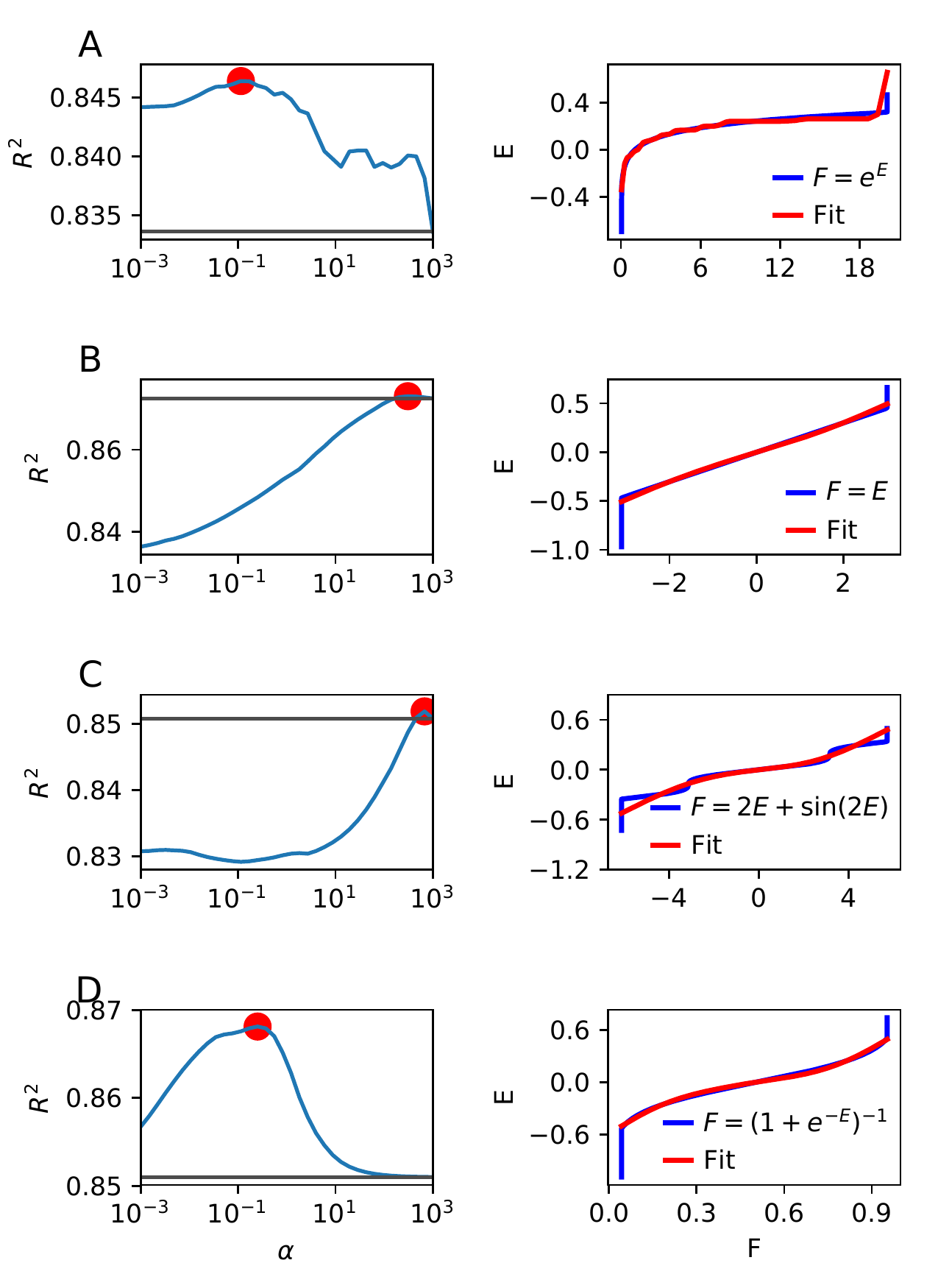}
\caption{Test of the inference method for learning the nonlinear scale on synthetic data.
A PWM was generated randomlly, with terms drawn from a normal distribution, and applied to the sequences from our dataset to simulate a ``true'' score ($E_{\rm{PWM}}$). Gaussian noise with 50\% of the standard deviation of $E_{\rm{PWM}}$ was added. The score was the converted into a free energy $F=f(E)$ (Methods), where $f$, the inverse function of $E$, was a ({\bf A}) linear, ({\bf B}) exponential, ({\bf C)} high frequency, or ({\bf D}) logistic transformation. The range of measured $F$ was cut off at the boundaries as in the data, resulting in the vertical lines in the right panels. Left panels: cross-validated fraction of explained variance, as a function of the regularization parameter $\alpha$ penalizing the second derivative of the function $E$. Large $\alpha$ implies very smooth functions, while small $\alpha$ allows for high-frequency variations.
These results the original PWM can be recovered as well as the non-linearity, except for its high-frequency components (C).}\label{fig:test_fit}
\end{figure*}

\begin{figure*}
\includegraphics[width=0.49\linewidth]{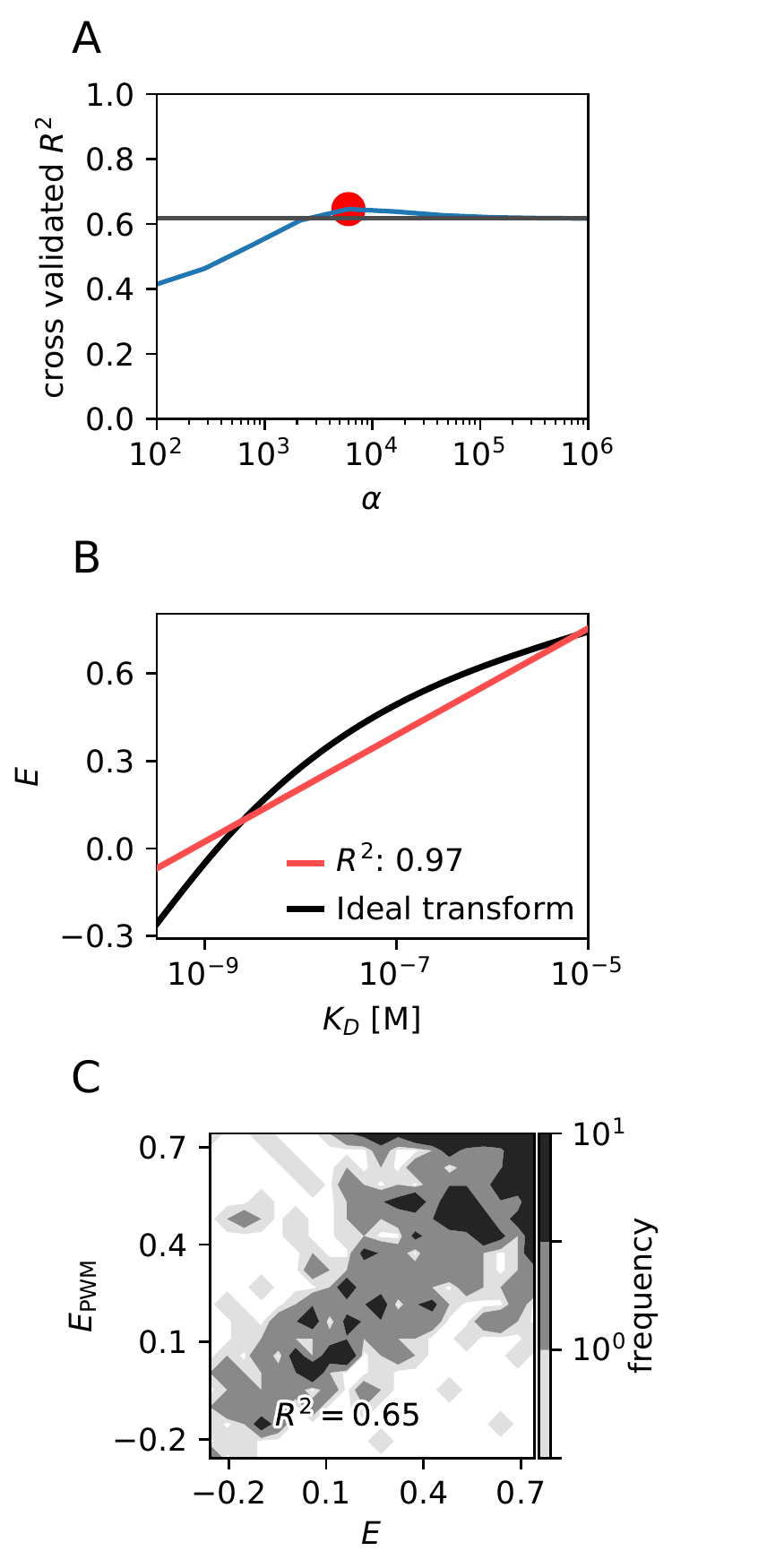}
\caption{Optimizing the non-linear scale $E$ for the PWM model on real data.
({\bf A}) Cross-validated fraction of explained variance, as a function of the regularization parameter $\alpha$ penalizing the second derivative of the function $E$.
({\bf B}) Optimized non-linear scale $E$ as a function of $F=\ln(K_D/c_0)$ (black), compared to identity (red). 
({\bf C}) Comparison between data and the PWM model with the optimal nonlinear scale.}\label{fig:best_energy_fun_boundary}
\end{figure*}

\begin{figure*}
\includegraphics[width=0.49\linewidth]{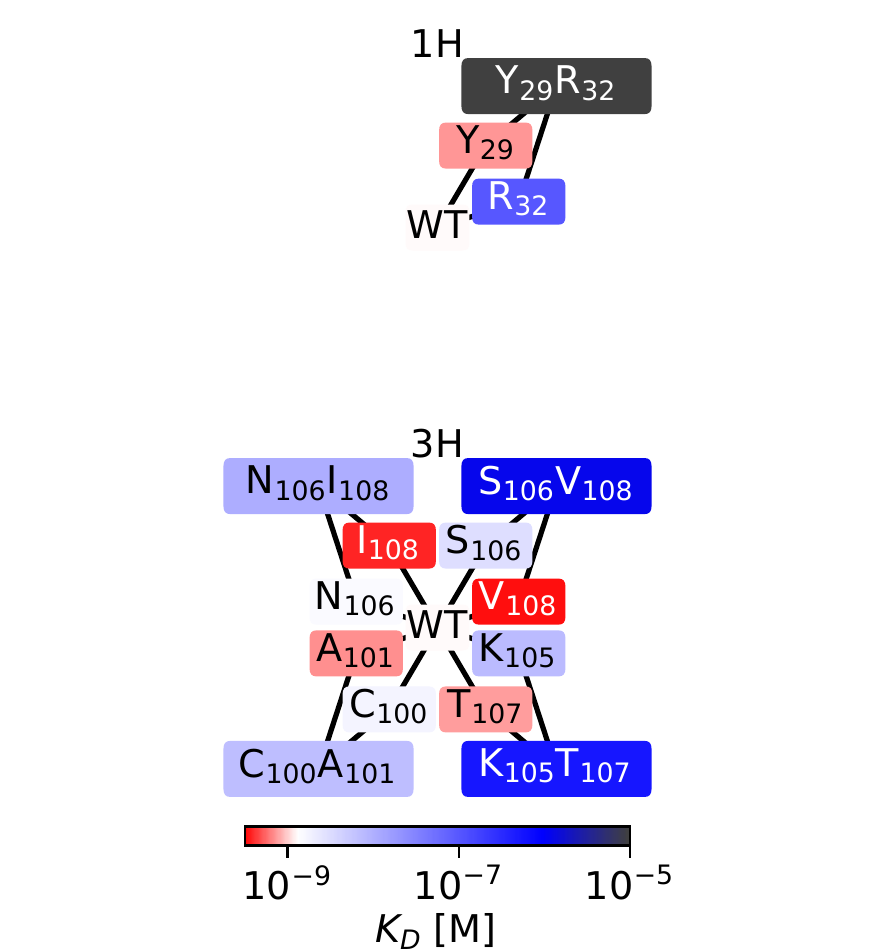}
\caption{All deleterious sign epistasis examples are shown for the 1H and 3H domains.}\label{fig:deleterious_sign_epistasis}
\end{figure*}

\begin{figure*}
\includegraphics[width=0.49\linewidth]{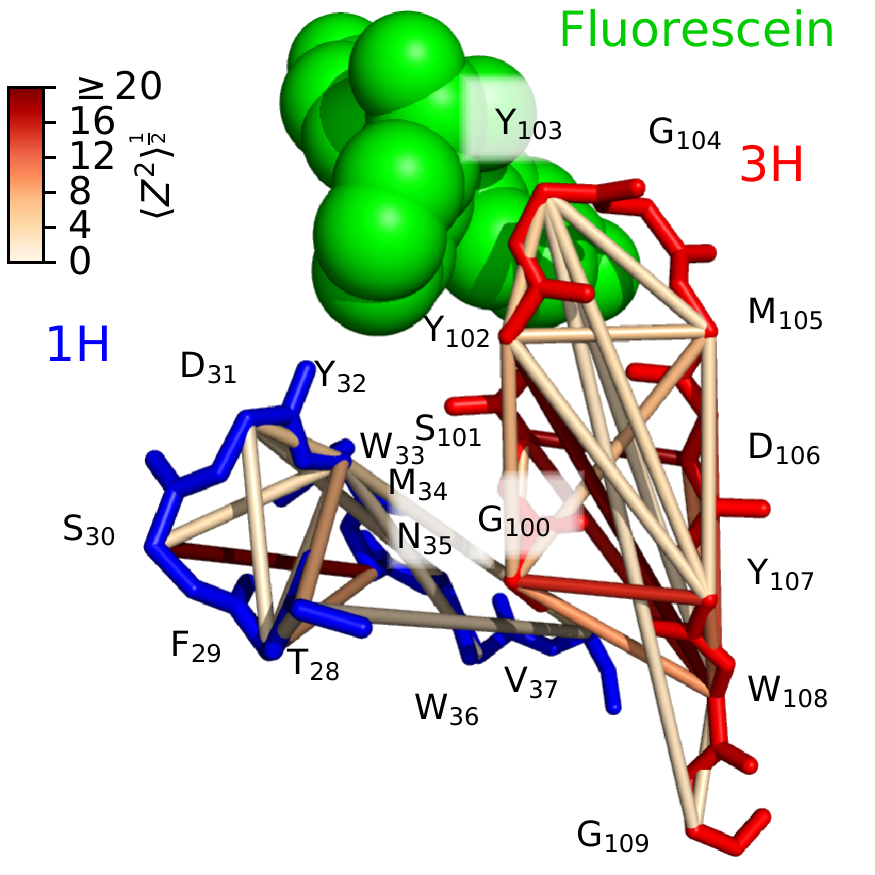}
\caption{Pairs of positions with standard epistatic effect ${\langle Z_{\rm{epi}}^2 \rangle }^{1/2}>3$ (with the mean taken over all measured amino-acid variants at the two positions) are superimposed on the wildtype antibody structure.}\label{fig:epistasis_structure}
\end{figure*}

\begin{figure*}
\includegraphics[width=0.65\linewidth]{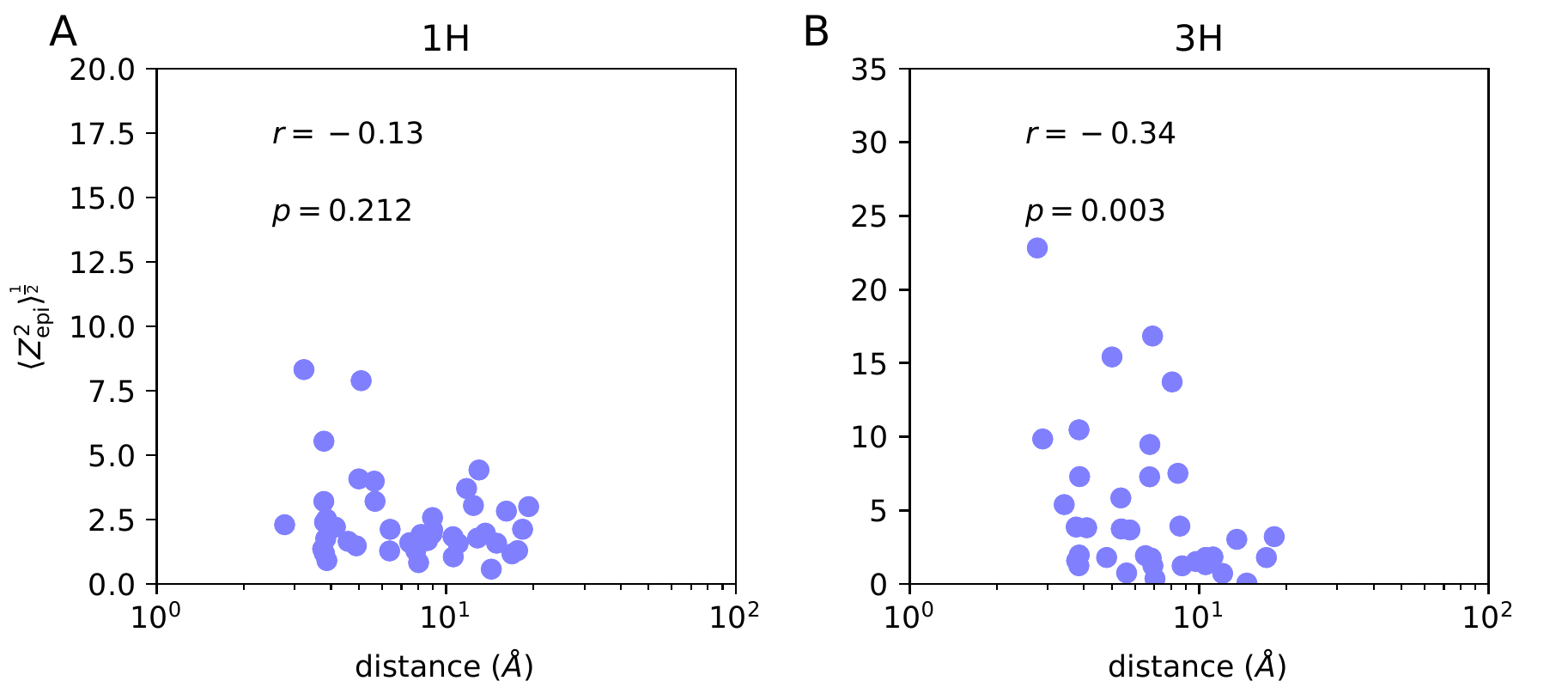}
\caption{Standard epistatic effect ${\langle Z_{\rm{epi}}^2 \rangle }^{1/2}$ versus distance between residues in the ({\bf A}) 1H and ({\bf B}) 3H domains.
}\label{fig:contact_v_epistasis}
\end{figure*}

\begin{figure*}
\includegraphics[width=0.65\linewidth]{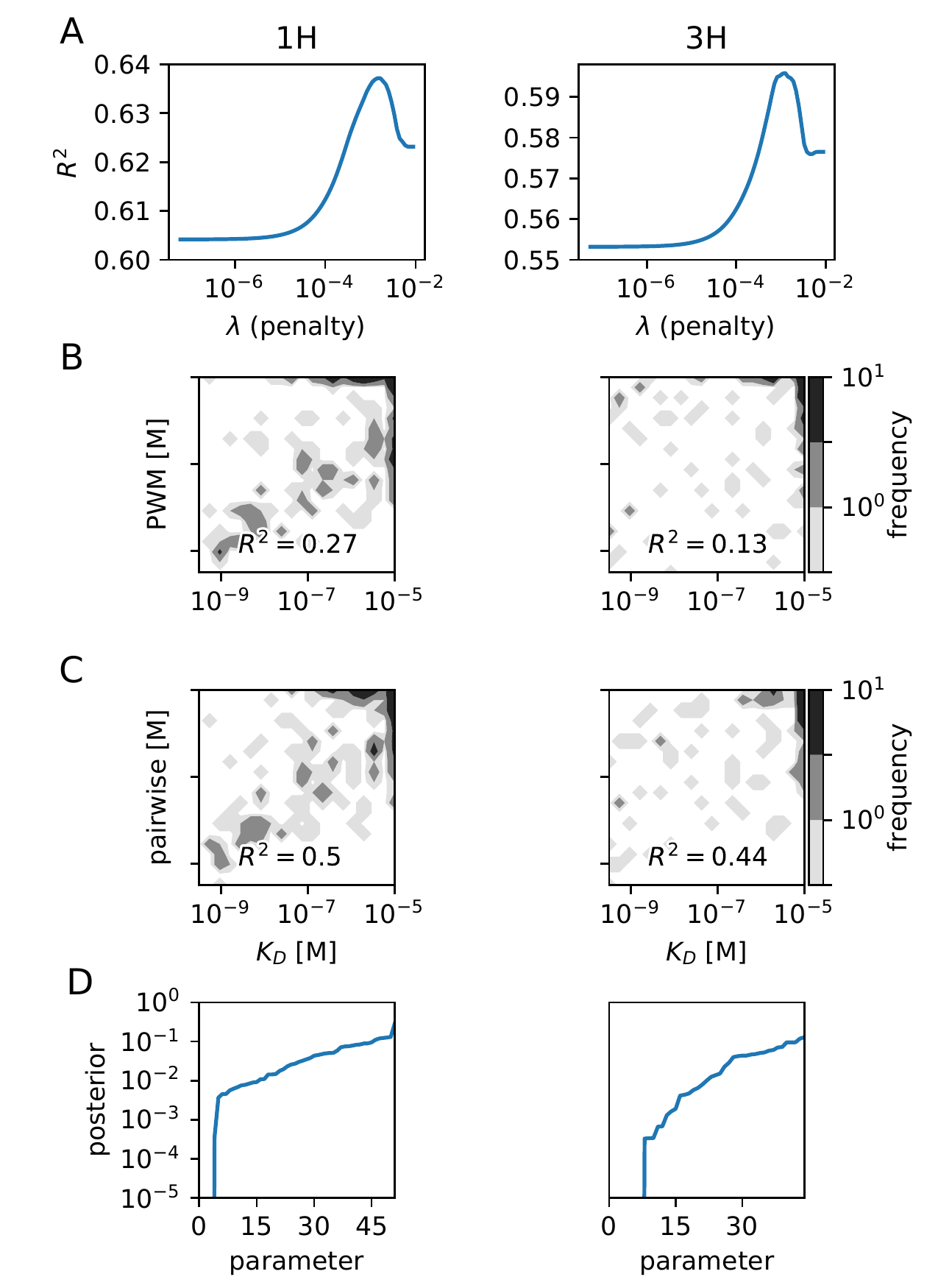}
\caption{Inferrence of the epistatic model. Left panels correspond to the 1H domain, and right panels to the 3H domain.
Parameters were fit by minimizing the mean squared model error with a L1 penalty on the parameters with coefficient $\lambda$. ({\bf A}) The cross-validated coefficient of determination ($1-{\rm standard\ error}^2/{\rm standard\ deviation}^2$) has a clear maximum as a function of $\lambda$. ({\bf B}) and ({\bf C}) Model prediction versus measurement of $F=\ln(K_D/c_0)$, for sequences involving a non-zero interacting term, when using ({\bf B}) the PWM model and ({\bf C}) the epistatic model with optimal $\lambda$.
({\bf D}) Rank-ordered posterior probability of $\tilde J$ parameters to be non-zero, according to the Bayesian Lasso posterior \cite{park_bayesian_2008}. Monte Carlo Markov Chains (MCMC) were used to estimate posterior probability of a parameter switching signs. Multiple (2 $\times$ number of non-zero parameters from Lasso optimization + 2) MCMCs were initialized (``thermalized'') for 227 steps $\times$ the number of parameters. After thermalization, the MCMCs were simulated for another 454 steps $\times$ the number of parameters. Parameter values were sampled at intervals where their autocorrelations were $\leq$ 0.1. The posterior probabilty was calculated as the fraction of time the parameter switched signs in the MCMCs.}\label{fig:bayesian_p_vals}
\end{figure*}

\end{document}